\documentclass[preprint,prd,aps,showpacs,showkeys,nofootinbib]{revtex4}
\usepackage{graphicx}
\begin{document}

\title{Neutron electric
dipole moment in CP violating BLMSSM}

\author{Shu-Min Zhao$^a$\footnote{email:smzhao@hbu.edu.cn}, Tai-Fu Feng$^{a}$\footnote{email:fengtf@hbu.edu.cn},
Ben Yan$^{a}$, Hai-Bin Zhang$^{a,b}$, \\Yin-Jie Zhang$^a$, Biao Chen$^a$, Xue-Qian Li$^c$}

\affiliation{$^a$Department of Physics, Hebei University, Baoding, 071002, China\\
$^b$Department of Physics, Dalian University of Technology,
Dalian, 116024, China\\
$^c$Department of Physics, Nankai University,
Tianjin, 300071, China}
\date{\today}

\begin{abstract}
Considering the CP violating phases,
we analyze the neutron electric dipole moment (EDM) in a CP violating supersymmetric extension
of the standard model where baryon
and lepton numbers are local gauge symmetries(BLMSSM). The contributions from
the one loop diagrams and the Weinberg operators are taken into account.
Adopting some assumptions on the relevant parameter space,
we give the numerical results analysis. The numerical results for the neutron EDM can reach $1.05\times 10^{-25}(e.cm)$,
which is about the experimental upper limit.
\end{abstract}



\maketitle

\section{Introduction}
Physicists have been interested in the minimal supersymmetric extension of the standard model
(MSSM) \cite{MSSM} for a long time. Considering the matter-antimatter asymmetry in the universe,
baryon number (B) should be broken. To explain the neutrino oscillation experiment, neutrino should
have tiny mass which can be induced by the seesaw mechanism from the heavy majorana neutrinos.
 Then lepton number (L) is also expected to be broken.
A minimal supersymmetric extension of
the SM with local gauged B and L(BLMSSM) is more favorite. With the assumption that B and L are
spontaneously broken gauge symmetries around ${\rm TeV}$ scale,
the author \cite{BLMSSM} examines two extensions of the SM.

B and L are spontaneously broken near the weak scale in BLMSSM. The light neutrinos
get mass from the seesaw mechanism, and proton decay is forbidden\cite{BLMSSM,BLMSSM1}. Therefore, a large desert between the
electroweak scale and grand unified scale is not necessary. This is the  main motivation for the BLMSSM.
Many possible signals of the MSSM at the LHC have been studied by the CMS\cite{CMS}  and ATLAS\cite{ATLAS}  experiments and they
set very strong bounds on the gluino and squarks masses with R-parity conservation. However, the predictions and bounds
for the collider experiments should be changed with the broken B and L symmetries\cite{BLMSSM,newresult}.

Flavour violation in the quark and leptonic sectors is suppressed naturally, because tree level flavor changing neutral currents
involving the quarks (charged leptons) are forbidden by the gauge symmetries and
particle content\cite{BLMSSM1}. We can look for lepton number
violation at the LHC, from the decays of right handed neutrinos\cite{BLMSSM1,rightneu}.
The baryon number violation in the decays of squarks
and gauginos\cite{BV} can also be detected at the LHC. For example, the baryon number violating decays of gluinos may
give rise to channels with multi-tops and multi-bottoms\cite{BLMSSM1}.

In BLMSSM, the mass and decays of the lightest CP-even Higgs
are investigated, where Yukawa couplings between Higgs doublets and exotic quarks are neglected\cite{BLMSSM,BLMSSM1}.
Taking into account the Yukawa couplings between Higgs and exotic quarks and ATLAS (CMS) results for
the neutral Higgs $(m_{h^0}\sim 124-126\;{\rm GeV})$\cite{Higgs},
the lightest CP-even Higgs decays $h^0\rightarrow\gamma\gamma$, $h\rightarrow ZZ (WW)$ are investigated
in BLMSSM\cite{weBLMSSM}.

As is known to all, the
magnetic dipole moments(MDMs)\cite{MDM,MDM1} and  electric dipole moments (EDMs)\cite{EDM} of fermions open a wide window
to find new physics beyond the SM.
The present experiment upper limit of the neutron EDM
is $1.1 \times10^{-25}\;e\cdot cm$ \cite{exp}. In SM, the CP phases of the
Cabibbo-Kobayashi-Maskawa (CKM) matrix elements fully induce very tiny neutron EDM,
and it is too tiny to be detected by experiment in the near future \cite{smt}.

In MSSM, many new
sources of the CP violation lead to large
contributions to the neutron EDM\cite{cp1,cp2} at one loop level, and the theoretical
prediction is larger than the present experiment upper bound. To consistent with the experiment
data, there are three approaches. The first is fine tuning, making the CP phases sufficiently small, i.e. $\le 10^{-2}$;
The second is that the supersymmetry spectra is very heavy at several TeV. Internal cancellation among the different
 contributions is the last one, and seems more reasonable\cite{xiangxiao}.

The theoretical analysis of EDMs for the fermions
 at one-loop level is well studied in MSSM.
 On other hand, authors obtain the corrections
to neutron EDM from the two-loop
Barr-Zee type diagrams involving the Higgs bosons\cite{barzee}.
For the neutron EDM, the two-loop gluino corrections and gluonic Weinberg operator induced by
two-loop gluino-squark diagrams are carried out in the literature\cite{neuEDM}.

In BLMSSM, there are many new CP violation sources and new particles.
Assuming that the Yukawa couplings between the Higgs and exotic quarks cannot be ignored,
we study the neutron EDM from the one loop diagrams and the Weinberg operator contributions in
this work. After this introduction, we briefly summarize the main ingredients
of the BLMSSM, and show the needed couplings for exotic quarks in section 2.
For the neutron EDM, the one loop diagrams' corrections and the Weinberg operator contributions
are presented in section 3. In section 4, we show the numerical results and study the relation between the
neutron EDM and the BLMSSM parameters. Our conclusions are summarized in section 5.

%%%%%%%%%%%%%%%%%%%%%%%%%%%%%%%%%%%%%%%%%%%%%%%%%%%%%%%%%%%%%%%%%%

\section{Some coupling in BLMSSM}

Enlarging the SM, the author obtains the BLMSSM \cite{BLMSSM} with the local gauge group of
$SU(3)_{_C}\otimes SU(2)_{_L}\otimes U(1)_{_Y}\otimes U(1)_{_B}\otimes U(1)_{_L}$.
In BLMSSM, to cancel L anomaly, there are the exotic superfields of the new leptons $\hat{L}_{_4}\sim(1,\;2,\;-1/2,\;0,\;L_{_4})$,
$\hat{E}_{_4}^c\sim(1,\;1,\;1,\;0,\;-L_{_4})$, $\hat{N}_{_4}^c\sim(1,\;1,\;0,\;0,\;-L_{_4})$,
$\hat{L}_{_5}^c\sim(1,\;2,\;1/2,\;0,\;-(3+L_{_4}))$, $\hat{E}_{_5}\sim(1,\;1,\;-1,\;0,\;3+L_{_4})$,
$\hat{N}_{_5}\sim(1,\;1,\;0,\;0,\;3+L_{_4})$. To break lepton number spontaneously, the superfields $\hat{\Phi}_{_L}\sim(1,\;1,\;0,\;0,\;-2)$ and
$\hat{\varphi}_{_L}\sim(1,\;1,\;0,\;0,\;2)$  are introduced, and they acquire nonzero vacuum expectation values (VEVs). Higgs mechanism
has a solid stone because of the detection of the lightest CP even Higgs at LHC\cite{Higgs}.

In the same way, the model includes the new quarks $\hat{Q}_{_4}\sim(3,\;2,\;1/6,\;B_{_4},\;0)$,
$\hat{U}_{_4}^c\sim(\bar{3},\;1,\;-2/3,\;-B_{_4},\;0)$,
$\hat{D}_{_4}^c\sim(\bar{3},\;1,\;1/3,\;-B_{_4},\;0)$,
$\hat{Q}_{_5}^c\sim(\bar{3},\;2,\;-1/6,\;-(1+B_{_4}),\;0)$, $\hat{U}_{_5}\sim(3,\;1,\;2/3,\;1+B_{_4},\;0)$,
$\hat{D}_{_5}\sim(3,\;1,\;-1/3,\;1+B_{_4},\;0)$, to cancel the B anomaly. $\hat{\Phi}_{_B}\sim(1,\;1,\;0,\;1,\;0)$ and
$\hat{\varphi}_{_B}\sim(1,\;1,\;0,\;-1,\;0)$ are the 'brand new' Higgs superfields. With nonzero vacuum expectation values (VEVs)
they break baryon number spontaneously. The exotic quarks are very heavy, because of the nonzero VEVs of $\Phi_{_B}$ and $\varphi_{_B}$.
Therefore, the BLMSSM also introduces the
superfields $\hat{X}\sim(1,\;1,\;0,\;2/3+B_{_4},\;0)$,
$\hat{X}^\prime\sim(1,\;1,\;0,\;-(2/3+B_{_4}),\;0)$ to make the unstable exotic quarks.
The lightest superfields X can be a candidate for dark matter.

In BLMSSM, the superpotential reads as follows
\begin{eqnarray}
&&{\cal W}_{_{BLMSSM}}={\cal W}_{_{MSSM}}+{\cal W}_{_B}+{\cal W}_{_L}+{\cal W}_{_X}\;,
\label{superpotential1}
\end{eqnarray}
with ${\cal W}_{_{MSSM}}$ representing the superpotential of the MSSM. The concrete forms of ${\cal W}_{_B},{\cal W}_{_L}$ and
${\cal W}_{_X}$ are
\begin{eqnarray}
&&{\cal W}_{_B}=\lambda_{_Q}\hat{Q}_{_4}\hat{Q}_{_5}^c\hat{\Phi}_{_B}+\lambda_{_U}\hat{U}_{_4}^c\hat{U}_{_5}
\hat{\varphi}_{_B}+\lambda_{_D}\hat{D}_{_4}^c\hat{D}_{_5}\hat{\varphi}_{_B}+\mu_{_B}\hat{\Phi}_{_B}\hat{\varphi}_{_B}
\nonumber\\
&&\hspace{1.2cm}
+Y_{_{u_4}}\hat{Q}_{_4}\hat{H}_{_u}\hat{U}_{_4}^c+Y_{_{d_4}}\hat{Q}_{_4}\hat{H}_{_d}\hat{D}_{_4}^c
+Y_{_{u_5}}\hat{Q}_{_5}^c\hat{H}_{_d}\hat{U}_{_5}+Y_{_{d_5}}\hat{Q}_{_5}^c\hat{H}_{_u}\hat{D}_{_5}\;,
\nonumber\\
%%%%%%%%%%%%%%%%%%%%%%%%%%%%%%%%%%%%%%%%%%%%%%%%%%%%%%%%%%%%%
&&{\cal W}_{_L}=Y_{_{e_4}}\hat{L}_{_4}\hat{H}_{_d}\hat{E}_{_4}^c+Y_{_{\nu_4}}\hat{L}_{_4}\hat{H}_{_u}\hat{N}_{_4}^c
+Y_{_{e_5}}\hat{L}_{_5}^c\hat{H}_{_u}\hat{E}_{_5}+Y_{_{\nu_5}}\hat{L}_{_5}^c\hat{H}_{_d}\hat{N}_{_5}
\nonumber\\
&&\hspace{1.2cm}
+Y_{_\nu}\hat{L}\hat{H}_{_u}\hat{N}^c+\lambda_{_{N^c}}\hat{N}^c\hat{N}^c\hat{\varphi}_{_L}
+\mu_{_L}\hat{\Phi}_{_L}\hat{\varphi}_{_L}\;,
\nonumber\\
%%%%%%%%%%%%%%%%%%%%%%%%%%%%%%%%%%%%%%%%%%%%%%%%%%%%%%%%%%%%%
&&{\cal W}_{_X}=\lambda_1\hat{Q}\hat{Q}_{_5}^c\hat{X}+\lambda_2\hat{U}^c\hat{U}_{_5}\hat{X}^\prime
+\lambda_3\hat{D}^c\hat{D}_{_5}\hat{X}^\prime+\mu_{_X}\hat{X}\hat{X}^\prime\;.
\label{superpotential-BL}
\end{eqnarray}
The soft breaking terms $\mathcal{L}_{_{soft}}$ of the BLMSSM can be found in our previous work\cite{weBLMSSM}.

To make the local gauge symmetry $SU(2)_{_L}\otimes U(1)_{_Y}\otimes U(1)_{_B}\otimes U(1)_{_L}$ broken
down to the electromagnetic symmetry $U(1)_{_e}$, the $SU(2)_L$ singlets $\Phi_{_B},\;\varphi_{_B},\;\Phi_{_L},\;
\varphi_{_L}$ and the $SU(2)_L$ doublets $H_{_u},\;H_{_d}$ should obtain nonzero VEVs
 $\upsilon_{_{B}},\;\overline{\upsilon}_{_{B}},\;\upsilon_{_L},\;\overline{\upsilon}_{_L}$
and $\upsilon_{_u},\;\upsilon_{_d}$ respectively.
\begin{eqnarray}
&&H_{_u}=\left(\begin{array}{c}H_{_u}^+\\{1\over\sqrt{2}}\Big(\upsilon_{_u}+H_{_u}^0+iP_{_u}^0\Big)\end{array}\right)\;,~~~~
H_{_d}=\left(\begin{array}{c}{1\over\sqrt{2}}\Big(\upsilon_{_d}+H_{_d}^0+iP_{_d}^0\Big)\\H_{_d}^-\end{array}\right)\;,
\nonumber\\
&&\Phi_{_B}={1\over\sqrt{2}}\Big(\upsilon_{_B}+\Phi_{_B}^0+iP_{_B}^0\Big)\;,~~~~~~~~~
\varphi_{_B}={1\over\sqrt{2}}\Big(\overline{\upsilon}_{_B}+\varphi_{_B}^0+i\overline{P}_{_B}^0\Big)\;,
\nonumber\\
&&\Phi_{_L}={1\over\sqrt{2}}\Big(\upsilon_{_L}+\Phi_{_L}^0+iP_{_L}^0\Big)\;,~~~~~~~~~~
\varphi_{_L}={1\over\sqrt{2}}\Big(\overline{\upsilon}_{_L}+\varphi_{_L}^0+i\overline{P}_{_L}^0\Big)\;,
\label{VEVs}
\end{eqnarray}

In Ref.\cite{weBLMSSM}, the mass matrixes of Higgs, exotic quarks and exotic scalar quarks are obtained.
In the mass basis, assuming CP conservation in exotic quark sector, we show the couplings between the neutral Higgs
and charged $(2/3,-1/3)$ exotic quarks. Here we investigate the CP violating effect and
modify our previous results.

To save space, we define $H_{_{\alpha}}^0(\alpha=1,2,3,\dots 8)=(h^0,H^0,A^0,G^0,h^0_{_B},H^0_{_B},A^0_{_B},G^0_{_B})$.
\begin{eqnarray}
&&{\cal L}_{_{H^0q' q'}}=\sum\limits_{\alpha=1}^8\sum\limits_{i,j=1}^2\Big\{
(\mathcal{N}^L_{_{H^0_{_{\alpha}}}})_{_{ij}}
H^0_{_{\alpha}}\overline{t}_{_{i+3}}P_{_L}t_{_{j+3}}
+(\mathcal{N}^R_{_{H^0_{_{\alpha}}}})_{_{ij}}
H^0_{_{\alpha}}\overline{t}_{_{i+3}}P_{_R}t_{_{j+3}}\nonumber\\&&\hspace{1.8cm}+(\mathcal{K}^L_{_{H^0_{_{\alpha}}}})_{_{ij}}
H^0_{_{\alpha}}\overline{b}_{_{i+3}}P_{_L}b_{_{j+3}}
+(\mathcal{K}^R_{_{H^0_{_{\alpha}}}})_{_{ij}}
H^0_{_{\alpha}}\overline{b}_{_{i+3}}P_{_R}b_{_{j+3}}\Big\},
\label{H0qq}
\end{eqnarray}
where $t_{_{i+3}},b_{_{i+3}}(i=1,2)$ are the exotic quark fields, the coupling contents
$(\mathcal{N}^L_{_{H^0_{_{\alpha}}}})_{_{ij}},
\;(\mathcal{N}^L_{_{H^0_{_{\alpha}}}})_{_{ij}},\;(\mathcal{N}^L_{_{H^0_{_{\alpha}}}})_{_{ij}},\;
(\mathcal{N}^L_{_{H^0_{_{\alpha}}}})_{_{ij}}$ are collected in the appendix.

The couplings between charged Higgs and exotic quarks are
\begin{eqnarray}
&&\mathcal{L}_{_{H^+ t^\prime b^\prime}}=\sum_{i,j=1}^2\Big[(\mathcal{N}^L_{_{H^+}})_{_{ij}} H^+\bar{t}_{_{i+3}}P_{_L}b_{_{j+3}}
+(\mathcal{N}^R_{_{H^+}})_{_{ij}} H^+\bar{t}_{_{i+3}}P_{_R}
b_{_{j+3}}\Big]\nonumber\\&&\hspace{1.8cm}+\sum_{i,j=1}^2\Big[(\mathcal{N}^L_{_{G^+}})_{_{ij}}G^+\bar{t}_{_{i+3}}P_{_L}b_{_{j+3}}
+(\mathcal{N}^R_{_{G^+}})_{_{ij}}G^+ \bar{t}_{_{i+3}}P_{_R}
b_{_{j+3}}\Big]+h.c.
\end{eqnarray}
with the coupling constants
\begin{eqnarray}
&&(\mathcal{N}^L_{_{H^+}})_{_{ij}}=-\Big(Y_{_{u_4}}(W_{_t}^{\dag})_{_{i2}}(U_{_b})_{_{1j}}+
Y_{_{d_5}}(W_{_t}^{\dag})_{_{i1}}(U_{_b})_{2j}\Big)\cos\beta,\nonumber\\&&
(\mathcal{N}^R_{_{H^+}})_{_{ij}}=\Big(Y_{_{d_4}}^*(U_{_t}^{\dag})_{_{i1}}(W_{_b})_{_{2j}}+
Y_{_{u_5}}^*(U_{_t}^{\dag})_{_{i2}}(W_{_b})_{_{1j}}\Big)\sin\beta,\nonumber\\&&
(\mathcal{N}^L_{_{G^+}})_{_{ij}}=-\Big(Y_{_{u_4}}(W_{_t}^{\dag})_{_{i2}}(U_{_b})_{_{1j}}+
Y_{_{d_5}}(W_{_t}^{\dag})_{_{i1}}(U_{_b})_{_{2j}}\Big)\sin\beta,\nonumber\\&&
(\mathcal{N}^R_{_{G^+}})_{_{ij}}=-\Big(Y_{_{d_4}}^*(U_{_t}^{\dag})_{_{i1}}(W_{_b})_{_{2j}}+
Y_{_{u_5}}^*(U_{_t}^{\dag})_{_{i2}}(W_{_b})_{_{1j}}\Big)\cos\beta.
\end{eqnarray}
Here, $\tan\beta=v_{u}/v_d$. $W_{_t}$ and $U_{_t}$ are the matrixes to diagonalize
the mass matrix of the up type exotic quarks. While, $W_{_b}$ and $U_{_b}$ are the matrixes to diagonalize
the mass matrix of the exotic quarks with charge $-1/3$.
 $A_{_{t_{i+3}}},\;A_{_{d_{i+3}}}(i=1,2)$ are the same with those in Ref.\cite{weBLMSSM}.
We also derive the couplings between photon (gluon) and exotic quarks.
\begin{eqnarray}
&&\mathcal{L}_{_{\gamma q'q'}}=-\frac{2e}{3}\sum_{i=1}^2\bar{t}_{_{i+3}}\gamma^{\mu}t_{_{i+3}}F_{\mu}
+\frac{e}{3}\sum_{i=1}^2\bar{b}_{_{i+3}}\gamma^{\mu}b_{_{i+3}}F_{\mu},\nonumber\\&&
\mathcal{L}_{_{g q'q'}}=-g_{_3}\sum_{i=1}^2\bar{t}_{_{i+3}}T^a\gamma^{\mu}t_{_{i+3}}G^a_{\mu}
-g_{_3}\sum_{i=1}^2\bar{b}_{_{i+3}}T^a\gamma^{\mu}b_{_{i+3}}G^a_{\mu},
\end{eqnarray}
where $F_{\mu}$ is electromagnetic field, $G^a_{\mu}$ is gluon field, $T^a\;(a=1,\;\cdots,\;8)$ denote the generators of
the strong $SU(3)$ gauge group.

 The couplings of gluino, exotic quark and exotic scalar quark have relation with Weinberg operator, and are written
 in the following form
\begin{eqnarray}
&&\mathcal{L}_{_{\Lambda t'\tilde{\mathcal{U}}}}=\sqrt{2}g_{_3}T^a\sum_{i=1}^4\sum_{j=1}^2\Big\{
(\mathcal{N}^L_{_{\tilde{\mathcal{U}}}})_{_{ij}}\tilde{\mathcal{U}}_{_i}^*\bar{\Lambda}_{_G}P_{_L}t_{_{j+3}}
+(\mathcal{N}^R_{\tilde{\mathcal{U}}})_{_{ij}}\tilde{\mathcal{U}}_{_i}^*\bar{\Lambda}_{_G}P_{_R}t_{_{j+3}}\Big\}+h.c.\nonumber\\&&
\mathcal{L}_{_{\Lambda b'\tilde{\mathcal{D}}}}=
\sqrt{2}g_{_3}T^a\sum_{i=1}^4\sum_{j=1}^2\Big\{
(\mathcal{N}^L_{_{\tilde{\mathcal{D}}}})_{_{ij}}\tilde{\mathcal{D}}_{_i}^*\bar{\Lambda}_{_G}P_{_L}b_{_{j+3}}
+(\mathcal{N}^R_{_{\tilde{\mathcal{D}}}})_{_{ij}}\tilde{\mathcal{D}}_{_i}^*\bar{\Lambda}_{_G}P_{_R}b_{_{j+3}}\Big\}+h.c.
\end{eqnarray}
where $\Lambda_{_G}$ is gluino field.  $\tilde{\mathcal{U}}_{_i},\tilde{\mathcal{D}}_{_i}(i=1,2,3,4)$ are exotic scalar quarks with charge $2/3$ and $-1/3$
respectively. The concrete forms of the coupling constants are
\begin{eqnarray}
%%%%%%%%%%%%%%%%%%%%%%%%%%%%%%%%%%%%%%%%%%%%%%%%%%%%%%%%%%%%%%%
&&(\mathcal{N}^L_{_{\tilde{\mathcal{U}}}})_{_{ij}}=U^{\dag}_{_{i4}}U^{_t}_{_{2j}}-U^{\dag}_{_{i1}}U^{_t}_{_{1j}},~~~~~~~~ (\mathcal{N}^R_{_{\tilde{\mathcal{U}}}})_{_{ij}}=U^{\dag}_{_{i2}}W^{_t}_{_{2j}}-U^{\dag}_{_{i3}}W^{_t}_{_{1j}},\nonumber\\&&
(\mathcal{N}^L_{_{\tilde{\mathcal{D}}}})_{_{ij}}=D^{\dag}_{_{i4}}U^{b}_{_{2j}}-D^{\dag}_{_{i1}}U^{b}_{_{1j}},~~~~~~~
(\mathcal{N}^R_{_{\tilde{\mathcal{D}}}})_{_{ij}}=D^{\dag}_{_{i2}}W^{b}_{_{2j}}-D^{\dag}_{_{i3}}W^{b}_{_{1j}}.
\end{eqnarray}

Using the scalar potential and the soft breaking terms, we write the mass squared matrix for superfields X.
\begin{equation}
-\mathcal{L}_{_X}=(X^*~~~X')\left(     \begin{array}{cc}
  |\mu_{_X}|^2+S_{_X} &-\mu_{_X}^*B_{_X}^* \\
    -\mu_{_X}B_{_X} & |\mu_{_X}|^2-S_{_X}\\
    \end{array}\right)  \left( \begin{array}{c}
  X \\  X'^*\\
    \end{array}\right),
   \end{equation}
    with $S_{_X}=\frac{g_{_B}^2}{2}(\frac{2}{3}+B_{_4})(v_{_B}^2-\bar{v}_{_B}^2)$.
  Adopting the unitary transformation,
   \begin{eqnarray}
   ~~~\left(     \begin{array}{c}
  X_{_1} \\  X_{_2}\\
    \end{array}\right) =Z_{_X}^{\dag}\left( \begin{array}{c}
  X \\  X'^*\\
    \end{array}\right),
   \end{eqnarray}
   the mass squared matrix for the superfields X are diagonalized.
   \begin{equation}
Z^{\dag}_{_X}\left(     \begin{array}{cc}
  |\mu_{_X}|^2+S_{_X} &-\mu_{_X}^*B_{_X}^* \\
    -\mu_{_X}B_{_X} & |\mu_{_X}|^2-S_{_X}\\
    \end{array}\right)  Z_{_X}=\left(     \begin{array}{cc}
 m_{_{X1}}^2 &0 \\
    0 & m_{_{X2}}^2\\
    \end{array}\right).
   \end{equation}
In mass basis, we obtain the couplings of quark-exotic quark and the superfields X.
    \begin{eqnarray}
  &&\mathcal{L}_{_{Xt'u}}=\sum_{i,j=1}^2\Big((\mathcal{N}^L_{_{t'}})_{_{ij}}X_{_j}\bar{t}_{_{i+3}}P_{_L}u^I
+(\mathcal{N}^R_{_{t'}})_{_{ij}}X_{_j}\bar{t}_{_{i+3}}P_{_R}u^I\Big)+h.c.\nonumber\\&&
\mathcal{L}_{_{Xb'd}}=\sum_{i,j=1}^2\Big((\mathcal{N}^L_{_{b'}})_{_{ij}}X_{_j}\bar{b}_{_{i+3}}P_{_L}d^I
+(\mathcal{N}^R_{_{b'}})_{_{ij}}X_{_j}\bar{b}_{_{i+3}}P_{_R}d^I\Big)+h.c.\label{LX}
\end{eqnarray}
where the coupling constants are
\begin{eqnarray}
%%%%%%%%%%%%%%%%%%%%%%%%%%%%%%%%%%%%%%%%%%
  &&(\mathcal{N}^L_{_{t'}})_{_{ij}}=-\lambda_{_1}(W_t^{\dag})_{_{i1}}(Z_{_X})_{_{1j}},~~~~~~~~ (\mathcal{N}^R_{_{t'}})_{_{ij}}=-\lambda_{_2}^*(U_{_t}^{\dag})_{_{i2}}(Z_{_X})_{_{2j}},\nonumber\\&&
  (\mathcal{N}^L_{_{b'}})_{_{ij}}=\lambda_{_1}(W_{_b}^{\dag})_{_{i1}}(Z_{_X})_{_{1j}},~~~~~~~~~~
  (\mathcal{N}^R_{_{b'}})_{_{ij}}=-\lambda_{_3}^*(U_{_b}^{\dag})_{_{i2}}(Z_{_X})_{_{2j}}. \label{couplingLX}
  \end{eqnarray}
%%%%%%%%%%%%%%%%%%%%%%%%%%%%%%%%%%%%%%%%%%%%%%%%%%%%%%%

\section{The corrections to the neutron EDM in BLMSSM }
The EDM $d_{_f}$ of the fermion can be obtained from the effective Lagrangian\cite{neuEDM}
\begin{eqnarray}
&&{\cal L}_{_{EDM}}=-{i\over2}d_{_f}\overline{f}\sigma^{\mu\nu}\gamma_5
fF_{_{\mu\nu}},
\label{eq1}
\end{eqnarray}
where $F_{_{\mu\nu}}$ represents the
electromagnetic field strength, $f$ is a fermion field.
This effective Lagrangian violates CP conservation obviously, and cannot be present
at tree level in the fundamental interactions. However in the CP violating electroweak theory,
it can be obtained at least at one loop level.

Because neutron is composed of quarks and gluons,
the chromoelectric dipole moment (CEDM)
$\overline{f}T^a\sigma^{\mu\nu}\gamma_5 fG^a_{_{\mu\nu}}$ of quarks can also give contributions
to the neutron EDM. $G^a_{_{\mu\nu}}$ represents the gluon field strength.
Weinberg has discovered a dimension-6 operator $f_{_{abc}}G_{_{\mu\rho}}^aG^{b\rho}_{_\nu}
G_{_{\lambda\sigma}}^c\epsilon^{\mu\nu\lambda\sigma}$ which violates CP conservation and it
 makes a very large contribution to the neutron EDM\cite{wenbergOP}.

The effective method is convenient to describe the CP violation
operators at loop level. At matching scale, we can integrate out the heavy particles in the loop
and obtain the resulting effective Lagrangian with a full set of CP violation
operators. Here, we just show the effective Lagrangian containing operators relevant to the neutron EDM.
\begin{eqnarray}
&&{\cal L}_{_{eff}}=\sum\limits_{i}^5C_{_i}(\Lambda){\cal O}_{_i}(\Lambda)\;,
\nonumber\\
&&{\cal O}_{_1}=\overline{q}\sigma^{\mu\nu}P_{_L}qF_{_{\mu\nu}}
\;,~~~~~{\cal O}_{_2}=\overline{q}\sigma^{\mu\nu}P_{_R}qF_{_{\mu\nu}}
\;,~~~{\cal O}_{_3}=\overline{q}T^a\sigma^{\mu\nu}P_{_L}qG^a_{_{\mu\nu}}
\;,\nonumber\\
&&{\cal O}_{_4}=\overline{q}T^a\sigma^{\mu\nu}P_{_R}qG^a_{_{\mu\nu}}
\;,~{\cal O}_{_5}=-{1\over6}f_{_{abc}}G_{_{\mu\rho}}^aG^{b\rho}_{_\nu}
G_{_{\lambda\sigma}}^c\epsilon^{\mu\nu\lambda\sigma}\;.
\label{eq3}
\end{eqnarray}
where $\Lambda$ is the energy scale, at which the Wilson coefficients $C_{_i}(\Lambda)$ are
evaluated.

\subsection{The corrections to neutron EDM in MSSM}
\begin{figure}[h]
\setlength{\unitlength}{1mm}
\centering
\includegraphics[width=4.0in]{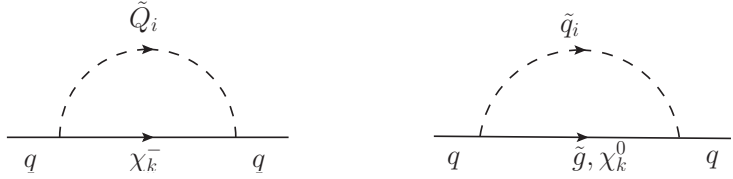}
\caption[]{The MSSM one loop self energy diagrams from which the corresponding triangle diagrams
are obtained by attaching a photon or a gluon in all possible ways.}\label{OLmssm}
\end{figure}

In SUSY, the one-loop supersymmetric corrections to the
quark EDMs and CEDMs are obtained\cite{neuEDM}. It can be divided into three types according to
the quark self-energy diagrams. The internal line particles are
 chargino and squark (neutralino and squark, gluino and squark).
The needed triangle diagrams are obtained by
attaching a photon or gluon on the internal lines of the
quark selfenergy diagrams in all possible ways. Calculating these diagrams one
can obtain the effective Lagrangian contributing to the
quark EDMs and CEDMs.
Before we
show the BLMSSM corrections, we present the one loop MSSM contributions to the quark EDMs and CEDMs, which
can be obtained in the previous works. The needed one loop self energy diagrams are shown in Fig.(\ref{OLmssm}).

 The one-loop neutralino-squark corrections to the
quark EDMs and CEDMs are
\begin{eqnarray}
&&d_{_{\chi_{_k}^0(1)}}^\gamma=e_{_q}{e\alpha\over16\pi s_{_{\rm w}}^2
c_{_{\rm w}}^2}\sum\limits_{i=1}^2\sum\limits_{k=1}^4{\bf Im}\Big((A_{_N}^q)_{_{k,i}}
(B_{_N}^q)^\dagger_{_{i,k}}\Big)
{m_{_{\chi_{_k}^ 0}}\over m_{_{\tilde{q}_i}}^2}B\Big({m_{_{\chi_{_k}^0}}^2
\over m_{_{\tilde{q}_i}}^2}\Big)
\;,\nonumber\\
&&d_{_{\chi_{_k}^0(1)}}^g={g_{_3}\alpha_{_s}\over64\pi s_{_{\rm w}}^2
c_{_{\rm w}}^2}\sum\limits_{i=1}^2\sum\limits_{k=1}^4{\bf Im}\Big((A_{_N}^q)_{_{k,i}}
(B_{_N}^q)^\dagger_{_{i,k}}\Big)
{m_{_{\chi_{_k}^0}}\over m_{_{\tilde{q}_i}}^2}B\Big({m_{_{\chi_{_k}^0}}^2
\over m_{_{\tilde{q}_i}}^2}\Big),
\label{eqa3-2}
\end{eqnarray}

Here $e_{_{u}}=2/3,\;e_{_{d}}=-1/3,\;\alpha=e^2/(4\pi),\;\alpha_{_s}=g_{_3}^2/(4\pi),\;s_{_W}=\sin\theta_{_W},\;c_{_W}=\cos\theta_{_W}$,
$\theta_{_W}$ is the Weinberg angle. $m_{_{\tilde{q}_i}}\;(i=1,2)$ are the scalar quarks masses and
$m_{_{\chi_{_k}^0}}\;(k=1,\;2,\;3,\;4)$ denote the eigenvalues of
neutralino mass matrix. We define the one loop function $B(r)$ as $B(r)=[2(r-1)^2]^{-1} [1+r+2r\ln
r/(r-1)]$. The concrete forms of the coupling constants $(A_{_N}^u)_{_{k,i}},\;(B_{_N}^u)_{_{k,i}},\;
(A_{_N}^d)_{_{k,i}},\;
(B_{_N}^d)_{_{k,i}}$ can be found in Ref.\cite{neuEDM}.

The contributions from the one loop gluino-squark diagrams are written as
\begin{eqnarray}
&&d_{_{\tilde{g}(1)}}^\gamma=-{2\over3\pi}e_{_q}e\alpha_{_s}\sum\limits
_{i=1}^2{\bf Im}\Big(({\cal Z}_{_{\tilde q}})_{_{2,i}}
({\cal Z}_{_{\tilde q}}^\dagger)_{_{i,1}}e^{-i\theta_{_3}}\Big)
{|m_{_{\tilde g}}|\over m_{_{\tilde{q}_i}}^2}B\Big({|m_{_{\tilde g}}|^2
\over m_{_{\tilde{q}_i}}^2}\Big)
\;,\nonumber\\
&&d_{_{\tilde{g}(1)}}^g={g_3\alpha_{_s}\over4\pi}\sum\limits
_{i=1}^2{\bf Im}\Big(({\cal Z}_{_{\tilde q}})_{_{2,i}} ({\cal
Z}_{_{\tilde q}}^\dagger)_{_{i,1}}e^{-i\theta_{_3}}\Big)
{|m_{_{\tilde g}}|\over m_{_{\tilde{q}_i}}^2}C\Big({|m_{_{\tilde g}}|^2
\over m_{_{\tilde{q}_i}}^2}\Big)\;,
\label{eqa3-1}
\end{eqnarray}
with $\theta_{_3}$ denoting the phase of
the gluino mass $m_{_{\tilde g}}$ and the loop function $C(r)=[6(r-1)^2]^{-1}[10r-26-(2r-18)\ln r/(r-1)]$. ${\cal
Z}_{_{\tilde q}}$ are the mixing matrices of the squarks, which diagonalize the
mass squared matrixes of squarks
${\cal Z}_{_{\tilde q}}^\dagger{\bf m}_{_{\tilde q}}^2{\cal
Z}_{_{\tilde q}} =diag(m_{_{{\tilde q}_1}}^2,\;m_{_{{\tilde
q}_2}}^2)$.

 We give the chargino-squark contributions here
\begin{eqnarray}
&&d_{_{\chi_{_k}^\pm(1)}}^\gamma={e\alpha\over4\pi s_{_{\rm w}}^2}
V_{_{qQ}}^\dagger V_{_{Qq}}
\sum\limits_{i,k}{\bf Im}\Big((A_{_C}^Q)_{_{k,i}}(B_{_C}^Q)^\dagger
_{_{i,k}}\Big){m_{_{\chi_{_k}^\pm}}\over m_{_{\tilde{Q}_i}}^2}
\nonumber\\
&&\hspace{1.2cm}\times
\Big[e_{_Q}B\Big({m_{_{\chi_{_k}^\pm}}^2\over m_{_{\tilde{Q}_i}}^2}\Big)
+(e_{_q}-e_{_Q})A\Big({m_{_{\chi_{_k}^\pm}}^2\over
m_{_{\tilde{Q}_i}}^2}\Big)\Big]
\;,\nonumber\\
&&d_{_{\chi_{_k}^\pm(1)}}^g={g_3\alpha\over4\pi s_{_{\rm w}}^2}
V_{_{qQ}}^\dagger V_{_{Qq}}
\sum\limits_{i,k}{\bf Im}\Big((A_{_C}^Q)_{_{k,i}}(B_{_C}^Q)^\dagger
_{_{i,k}}\Big)
{m_{_{\chi_{_k}^\pm}}\over m_{_{\tilde{Q}_i}}^2}
B\Big({m_{_{\chi_{_k}^\pm}}^2\over m_{_{\tilde{Q}_i}}^2}\Big)\;,
\label{eqa3-4}
\end{eqnarray}
where $m_{_{\chi_{_k}^\pm}}\;(k=1,\;2)$
denote the chargino masses, $V$ is the CKM matrix, and the
concrete form of $A(r)$ is
$A(r)=[2(1-r)^{2}]^{-1}[3-r+2\ln r/(1-r)]$.
One can obtain the concrete forms of
$(A_{_C}^d)_{_{k,i}},\;(B_{_C}^d)_{_{k,i}}
,\;(A_{_C}^u)_{_{k,i}}
,\;(B_{_C}^u)_{_{k,i}}$ in the literature\cite{neuEDM}.

\begin{figure}[h]
\setlength{\unitlength}{1mm}
\centering
\includegraphics[width=1.8in]{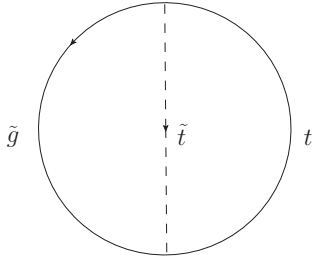}
\caption[]{The MSSM two loop diagram contributing to the Weinberg operator
 by attaching three gluons in all possible ways.}\label{Wmssm}
\end{figure}

In MSSM, the two-loop "gluino-squark" diagram shown in Fig.(\ref{Wmssm}) gives contributions
to the gluonic Weinberg operator, whose Wilson coefficient is \cite{neuEDM,wenbergOP}
\begin{eqnarray}
&&C^{SUSY}_{_5}=-3\alpha_{_s}m_{_t}\Big({g_3\over4\pi}\Big)^3
{\bf Im}\Big(({\cal Z}_{_{\tilde t}})_{_{2,2}}
({\cal Z}_{_{\tilde t}})^\dagger_{_{2,1}}\Big)
{m_{_{\tilde{t}_1}}^2-m_{_{\tilde{t}_2}}^2\over |m_{_{\tilde g}}|^5}
H({m_{_{\tilde{t}_1}}^2\over |m_{_{\tilde g}}|^2},{m_{_{\tilde{t}_2}}^2
\over |m_{_{\tilde g}}|^2},{m_{_t}^2\over |m_{_{\tilde g}}|^2})
\label{eq13}
\end{eqnarray}
Here, the $H$ function can be found in \cite{wenbergOP}. ${\cal Z}_{_{\tilde t}}$ is the matrix to diagonalize the
mass squared matrix of stop. Up to now, in MSSM the dominant contributions to the neutron EDM are complete.
\subsection{The corrections to the neutron EDM from superfields in BLMSSM}
\begin{figure}[h]
\setlength{\unitlength}{1mm}
\centering
\includegraphics[width=2.8in]{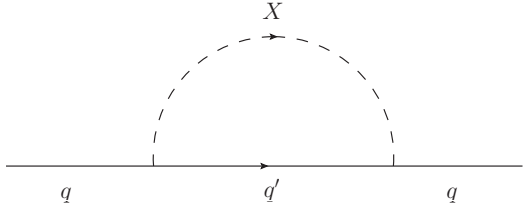}
\caption[]{The one loop selfenergy diagram from which the corresponding triangle diagrams
are obtained by attaching a photon or a gluon in all possible ways.}\label{oneloop}
\end{figure}

In BLMSSM, the quark, exotic quark and superfields X have a coupling at tree level in Eq.(\ref{LX})
to avoid stability for the exotic quarks.
Therefore, there are one loop triangle diagrams obtained by attaching a photon or a gluon in all possible ways on the self energy
 diagram in Fig.(\ref{oneloop}). Then we get the quark EDMs and CEDMs.
\begin{eqnarray}
&&d^{\gamma}_{_{X_j}}=-\frac{e_{_q}e}{64 \pi
   ^2}\sum_{i,j=1}^2\frac{m_{_{q_{_{i+3}}}}}{m_{_{X_{_{j}}}}^2}{\bf Im}\Big((\mathcal{N}^L_{_{q'}})_{_{ij}}(\mathcal{N}^R_{_{q'}})^{\dag}_{_{ji}}\Big)
   A\Big(\frac{m_{_{q_{_{i+3}}}}^2}{m_{_{X_{_{j}}}}^2}\Big),
\nonumber\\&&
d^{g}_{_{X_j}}=-\frac{g_{_3}}{64 \pi
   ^2}\sum_{i,j=1}^2\frac{m_{_{q_{_{i+3}}}}}{m_{_{X_{_{j}}}}^2}{\bf Im}\Big((\mathcal{N}^L_{_{q'}})_{_{ij}}(\mathcal{N}^R_{_{q'}})^{\dag}_{_{ji}}\Big)
   A\Big(\frac{m_{_{q_{_{i+3}}}}^2}{m_{_{X_{_{j}}}}^2}\Big),
\end{eqnarray}
where $q=t(b),q'=t'(b')$, $e_{_{t^{'}}}=2/3, e_{_{b^{'}}}=-1/3$, $m_{_{X_{_{j}}}}(j=1,\;2)$ denote the masses of the superfields X, $m_{_{t_{_{i+3}}}}(i=1,2)$ represent the masses
of the up type exotic quarks,  $m_{_{b_{_{i+3}}}}(i=1,2)$ are the masses
of the down type exotic quarks.

\begin{figure}[h]
\setlength{\unitlength}{1mm}
\centering
\includegraphics[width=3.0in]{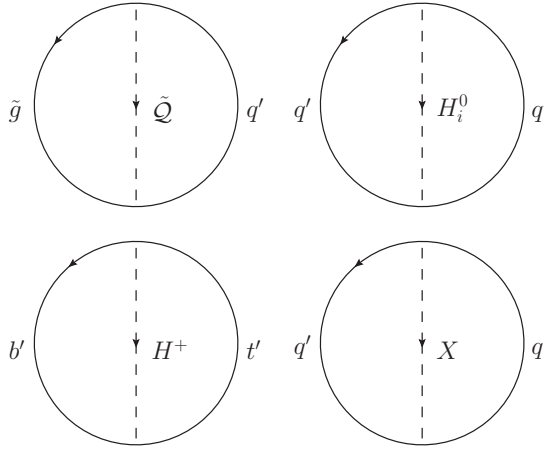}
\caption[]{The BLMSSM two loop diagrams contributing to the Weinberg operator
 by attaching three gluons in all possible ways.}\label{Wblmssm}
\end{figure}

The Weinberg operator comes from the two loop gluon diagrams. In this model, there are four exotic quarks and eight exotic scalar quarks,
which leads to a lot of diagrams contributing to the Wilson coefficient $C_{_5}(\Lambda)$ of the Weinberg operator. These two loop diagrams are drawn in Fig.(\ref{Wblmssm}).
The two loop "gluino-exotic quark-exotic scalar quark" diagrams have corrections
to the Wilson coefficient of the purely gluonic Weinberg operator, and the
results are presented here.
\begin{eqnarray}
&&C_{_5}^{\Lambda t'\tilde{\mathcal{U}}}=3\alpha_{_s}\Big({g_{_3}\over4\pi}\Big)^3\sum_{i=1}^2m_{_{t_{i+3}}}\Big[
{\bf Im}\Big((\mathcal{N}^L_{_{\tilde{\mathcal{U}}}})^{\dag}_{_{i2}}(\mathcal{N}^R_{_{\tilde{\mathcal{U}}}})_{_{2i}}\Big)
{m_{_{\tilde \mathcal{U}_1}}^2-m_{_{\tilde \mathcal{U}_2}}^2\over |m_{_{\tilde g}}|^5}
H({m_{_{\tilde \mathcal{U}_1}}^2\over |m_{_{\tilde g}}|^2},{m_{_{\tilde \mathcal{U}_2}}^2
\over |m_{_{\tilde g}}|^2},{m_{_{t_{i+3}}}^2\over |m_{_{\tilde g}}|^2})\nonumber\\&&
%%%%%%%%%%%%%%%%%%%%%%%%%%%%%%%%%%%%%%%%%%%%%%%%%%%%%%%%%%%%%%%%%%%
\hspace{1.8cm}+{\bf Im}\Big((\mathcal{N}^L_{_{\tilde{\mathcal{U}}}})^{\dag}_{_{i4}}(\mathcal{N}^R_{_{\tilde{\mathcal{U}}}})_{_{4i}}\Big)
{m_{_{\tilde \mathcal{U}_3}}^2-m_{_{\tilde \mathcal{U}_4}}^2\over |m_{_{\tilde g}}|^5}
H({m_{_{\tilde \mathcal{U}_3}}^2\over |m_{_{\tilde g}}|^2},{m_{_{\tilde \mathcal{U}_4}}^2
\over |m_{_{\tilde g}}|^2},{m_{_{t_{i+3}}}^2\over |m_{_{\tilde g}}|^2})\Big],
\nonumber\\
&&C_{_5}^{\Lambda b'\tilde{\mathcal{D}}}=3\alpha_{_s}\Big({g_{_3}\over4\pi}\Big)^3\sum_{i=1}^2m_{_{b_{i+3}}}\Big[
{\bf Im}\Big((\mathcal{N}^L_{_{\tilde{\mathcal{D}}}})^{\dag}_{_{i2}}(\mathcal{N}^R_{_{\tilde{\mathcal{D}}}})_{_{2i}}\Big)
{m_{_{\tilde \mathcal{D}_1}}^2-m_{_{\tilde \mathcal{D}_2}}^2\over |m_{_{\tilde g}}|^5}
H({m_{_{\tilde \mathcal{D}_1}}^2\over |m_{_{\tilde g}}|^2},{m_{_{\tilde \mathcal{D}_2}}^2
\over |m_{_{\tilde g}}|^2},{m_{_{b_{i+3}}}^2\over |m_{_{\tilde g}}|^2})\nonumber\\&&
%%%%%%%%%%%%%%%%%%%%%%%%%%%%%%%%%%%%%%%%%%%%%%%%%%%%%%%%%%%%%%%%%%%
\hspace{1.8cm}+{\bf Im}\Big((\mathcal{N}^L_{_{\tilde{\mathcal{D}}}})^{\dag}_{_{i4}}(\mathcal{N}^R_{_{\tilde{\mathcal{D}}}})_{_{4i}}\Big)
{m_{_{\tilde \mathcal{D}_3}}^2-m_{_{\tilde \mathcal{D}_4}}^2\over |m_{_{\tilde g}}|^5}
H({m_{_{\tilde \mathcal{D}_3}}^2\over |m_{_{\tilde g}}|^2},{m_{_{\tilde \mathcal{D}_4}}^2
\over |m_{_{\tilde g}}|^2},{m_{_{b_{i+3}}}^2\over |m_{_{\tilde g}}|^2})\Big].
\end{eqnarray}
Here, $m_{_{\tilde \mathcal{U}_i}}(i=1,2,3,4)$ denote the masses of the up type exotic scalar quarks,
 and $m_{_{\tilde \mathcal{D}_i}}(i=1,2,3,4)$ are the masses of the down type exotic scalar quarks.

 From the couplings of neutral Higgs and exotic quarks in Eq.(\ref{H0qq}), we obtain their
 contributions to the Wilson coefficient $C_{_5}(\Lambda)$.
\begin{eqnarray}
&&C_{_5}^{H^0t't'}=2g_{_3}^3\Big({1\over4\pi}\Big)^4\sum_{\alpha=1}^{8}\Big[\sum_{i=1}^2{\bf Im}\Big((\mathcal{N}^R_{_{H^0_{_{\alpha}}}})_{_{ii}}(\mathcal{N}^L_{_{H^0_{_{\alpha}}}})^{\dag}_{_{ii}}\Big)\frac{1}{m_{_{t_{i+3}}}^2}
h(m_{_{t_{i+3}}},m_{_{H^0_{_{\alpha}}}})\nonumber\\&&
\hspace{1.8cm}+{\bf Im}\Big((\mathcal{N}^R_{_{H^0_{_{\alpha}}}})_{_{12}}(\mathcal{N}^L_{_{H^0_{_{\alpha}}}})^{\dag}_{_{21}}\Big)\frac{1}{m_{_{t_{4}}}m_{_{t_{5}}}}
h'(m_{_{t_{4}}},m_{_{t_{5}}},m_{_{H^0_{_{\alpha}}}})\Big],
\nonumber\\
&&C_{_5}^{H^0b'b'}=2g_{_3}^3\Big({1\over4\pi}\Big)^4\sum_{\alpha=1}^{8}\Big[\sum_{i=1}^2{\bf Im}\Big((\mathcal{K}^R_{_{H^0_{_{\alpha}}}})_{_{ii}}(\mathcal{K}^L_{_{H^0_{_{\alpha}}}})^{\dag}_{_{ii}}\Big)\frac{1}{m_{_{b_{i+3}}}^2}
h(m_{_{b_{i+3}}},m_{_{H^0_{_{\alpha}}}})\nonumber\\&&
\hspace{1.8cm}+{\bf Im}\Big((\mathcal{K}^R_{_{H^0_{_{\alpha}}}})_{_{12}}(\mathcal{K}^L_{_{H^0_{_{\alpha}}}})^{\dag}_{_{21}}\Big)\frac{1}{m_{_{b_{4}}}m_{_{b_{5}}}}
h'(m_{_{b_{4}}},m_{_{b_{5}}},m_{_{H^0_{_{\alpha}}}})\Big],
\end{eqnarray}
where the functions $h(x,y)$ and $h'(x,y,z)$ are given in \cite{wenbergOP}.
 In the same way, the contributions from charged Higgs and exotic quarks are written in the following form.
\begin{eqnarray}
&&C_{_5}^{H^+t'b'}=2g_{_3}^3\Big({1\over4\pi}\Big)^4\sum_{i,j=1}^{2}\frac{1}{m_{_{t_{i+3}}}m_{_{b_{j+3}}}}\Big[{\bf Im}\Big((\mathcal{N}^R_{_{H^+}})_{_{ij}}(\mathcal{N}^L_{_{H^+}})^{\dag}_{_{ji}}\Big)
h'(m_{_{t_{i+3}}},m_{_{b_{j+3}}},m_{_{H^+}})\nonumber\\&&
\hspace{1.8cm}+{\bf Im}\Big((\mathcal{N}^R_{_{G^+}})_{_{ij}}(\mathcal{N}^L_{_{G^+}})^{\dag}_{_{ji}}\Big)h'(m_{_{t_{i+3}}},m_{_{b_{j+3}}},m_{_{G^+}})\Big].
\end{eqnarray}

The coupling of quark-exotic quark-superfield X in Eq.(\ref{LX}) can not only give the corrections to the neutron EDM at
one loop level, but also contribute to the Wilson coefficient $C_{_5}(\Lambda)$ at two loop level.
\begin{eqnarray}
&&C_{_5}^{Xq'q}=2g_{_3}^3\Big({1\over4\pi}\Big)^4\sum_{i,j=1}^{2}\Big[{\bf Im}\Big((\mathcal{N}^R_{t'})_{_{ij}}(\mathcal{N}^L_{t'})^{\dag}_{_{ji}}\Big)\frac{1}{m_{_{t_{i+3}}}m_{_{t}}}h'(m_{_{t_{i+3}}},m_{_{t}},m_{_{X_j}})\nonumber\\&&
\hspace{1.8cm}+{\bf Im}\Big((\mathcal{N}^R_{b'})_{_{ij}}(\mathcal{N}^L_{b'})^{\dag}_{_{ji}}\Big)\frac{1}{m_{_{b_{i+3}}}m_{_{b}}}h'(m_{_{b_{i+3}}},m_{_{b}},m_{_{X_j}})\Big].
\end{eqnarray}

The results obtained at the matching scale $\Lambda$ should be transformed down to
the chirality breaking scale $\Lambda_{_\chi}$ \cite{rge}. So the renormalization group equations
(RGEs) for the Wilson
coefficients of the Weinberg operator and the quark EDMs, CEDMs should be solved.
The relations between the results at two deferent scales $\Lambda$ and $\Lambda_{_\chi}$
are presented here.
\begin{eqnarray}
d_{_q}^\gamma(\Lambda_{_\chi})=1.53d_{_q}^\gamma(\Lambda)
,~~~~ d_{_q}^g(\Lambda_{_\chi})=3.4d_{_q}^g(\Lambda)
,~~~~C_{_5}(\Lambda_{_\chi})=3.4C_{_5}(\Lambda)\;,
\label{eq14}
\end{eqnarray}
There are three type contributions to the quark EDM, 1 the
electroweak contribution $d_{_q}^\gamma$, 2 the CDEM contribution $d_{_q}^g$,
3 the Weinberg operator contribution. Each type contribution can be calculated numerically.
At a low scale, the quark EDM can be obtained from $d_{_q}^\gamma, d_{_q}^g$ and $C_{_5}(\Lambda_{_\chi})$ by
the following formula\cite{nda}.
\begin{eqnarray}
&&d_{_q}=d_{_q}^\gamma+{e\over4\pi}d_{_q}^g+{e\Lambda_{_\chi}
\over4\pi}C_{_5}(\Lambda_{_\chi})\;.
\label{eq15}
\end{eqnarray}

From the quark model, the EDM of neutron is derived from u quark's EDM $d_{_u}$ and d
quark's EDM $d_{_d}$ with the following expression
\begin{eqnarray}
&&d_{_n}={1\over3}(4d_{_d}-d_{_u}).
\label{eq16}
\end{eqnarray}

\section{The numerical analysis \label{sec3}}
The lightest neutral CP even Higgs is discovered by LHC,
with a mass $m_{_{h_0}}\simeq125.9\;{\rm GeV}$, which gives
the most stringent constraint on the parameter space. Furthermore,
the ATLAS and CMS experiments announce
the Higgs production and decay into diphoton channel are larger than
SM predictions with a factor $1.2\sim2$.
The current values of the ratios are \cite{Higgs,Htovv}:
$R_{\gamma\gamma}=1.6\pm0.4\;,
R_{VV^*}=0.94\pm0.40\;,(V=Z,\;W).$ CMS collaboration declare that it is impossible to find a SM Higgs with mass from 127.5 to 600 GeV. Considering all the above
constraints, we obtain a parameter space in our recent work\cite{weBLMSSM}.

In this work to obtain the numerical results, the relevant parameters in SM are shown here.
\begin{eqnarray}
&&\alpha_s(m_{_{\rm Z}})=0.118,\;\;\alpha(m_{_{\rm Z}})=1/128,
\;\;s_{_{\rm W}}^2(m_{_{\rm Z}})=0.23,\nonumber\\
&&m_{_{\rm W}}=80.4\;{\rm GeV},\;
m_t=174.2\;{\rm GeV},\;\;m_b=4.2\;{\rm GeV},\nonumber\\&&
m_u=2.3\times10^{-3}\;{\rm GeV},~~~~~\;m_d=4.8\times10^{-3}\;{\rm GeV},
\label{PDG-SM}
\end{eqnarray}
To simplify the numerical discussion, we suppose $A_{_t},A_{_b}, A_{_{\nu_4}},A_{_{e_4}},A_{_{\nu_5}},
A_{_{\nu_5}},A_{_{u_4}},A_{_{u_5}},A_{_{d_4}},A_{_{d_5}}$, $A_{_{BQ}},A_{_{BU}},A_{_{BD}}$ and $Y_{_{u_4}},Y_{_{u_5}},Y_{_{d_4}},Y_{_{d_5}}$
are all real parameters.

 The other parameters are adopted\cite{weBLMSSM}.
\begin{eqnarray}
&&\tan\beta=\tan\beta_{_B}=\tan\beta_{_L}=2,\;~~B_4=L_4={3\over2},\;
\nonumber\\
&&m_{_{\tilde{Q}_3}}=m_{_{\tilde{U}_3}}=m_{_{\tilde{D}_3}}=1\;{\rm TeV},\;~~m_{_{Z_B}}=m_{_{Z_L}}=1\;{\rm TeV},
\nonumber\\
&&m_{_{\tilde{U}_4}}=m_{_{\tilde{D}_4}}=m_{_{\tilde{Q}_5}}=m_{_{\tilde{U}_5}}
=m_{_{\tilde{D}_5}}=1\;{\rm TeV},\;\; m_{_{\tilde{Q}_4}}=790\;{\rm GeV},\nonumber\\
&&m_{_{\tilde{L}_4}}=m_{_{\tilde{\nu}_4}}=m_{_{\tilde{E}_4}}=m_{_{\tilde{L}_5}}=m_{_{\tilde{\nu}_5}}
=m_{_{\tilde{E}_5}}=1\;{\rm TeV}\;,\nonumber\\
&&A_{_{\nu_4}}=A_{_{e_4}}=A_{_{\nu_5}}=A_{_{\nu_5}}=A_{_{u_4}}=A_{_{u_5}}=A_{_{d_4}}=A_{_{d_5}}=550\;{\rm GeV}
\;,\nonumber\\
&&\upsilon_{_{B_t}}=\sqrt{\upsilon_{_B}^2+\overline{\upsilon}_{_B}^2}=3\;{\rm TeV}\;,\;\;
\upsilon_{_{L_t}}=\sqrt{\upsilon_{_L}^2+\overline{\upsilon}_{_L}^2}=3\;{\rm TeV}\;,
\nonumber\\
&&A_{_{BQ}}=A_{_{BU}}=A_{_{BD}}=-A_{_b}=-A_{_t}=1\;{\rm TeV},~~m_{_1}=\;{\rm TeV},
\nonumber\\
&&Y_{_{u_4}}=0.76Y_t,\;Y_{_{d_4}}=0.7Y_b,\;Y_{_{u_5}}=0.7Y_b,\;Y_{_{d_5}}=0.13Y_t,
\nonumber\\
&&m_2=750\;{\rm GeV}\;,\;\mu_{_B}=500\;{\rm GeV},\;\Lambda=1\;{\rm TeV},\;\Lambda_{\chi}=1\;{\rm GeV},
\nonumber\\&&\lambda_{_Q}=\lambda_{_u}=\lambda_{_d}=0.5,\;~~|m_{\tilde{g}}|=300\;{\rm GeV},~~B_{_B}=1\;{\rm TeV},
\nonumber\\&&m_{_{\nu_4}}=m_{_{\nu_5}}=90\;{\rm GeV},\;\;m_{_{e_4}}=m_{_{e_5}}=100\;{\rm GeV},~\;m_{_{\mu}}=-800\;{\rm GeV}.
\label{assumption1}
\end{eqnarray}

\begin{figure}[h]
\setlength{\unitlength}{1mm}
\centering
\includegraphics[width=4.0in]{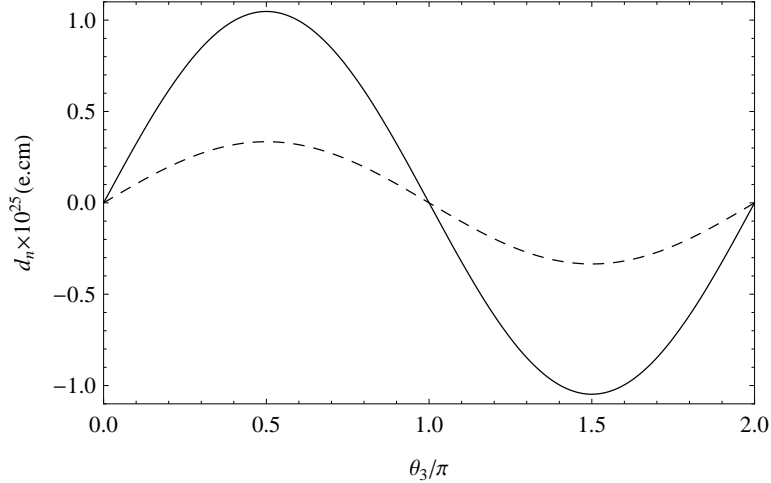}
\caption[]{As $B_{_X}=300\;{\rm GeV},\;\lambda_{_1}=\lambda_{_2}=\lambda_{_3}=0.1,\;|\mu_{_X}|=2400\;{\rm GeV},\;\theta_{_X}=0$
and $m_{_{\tilde{Q}_1}}=m_{_{\tilde{U}_1}}=m_{_{\tilde{D}_1}}=MQ$, the solid-line and dashed-line represent $d_n$ varying with $\theta_{_3}$, for
$MQ=1000{\rm GeV}$ and $1200{\rm GeV}$ respectively.}\label{theta3}
\end{figure}

The CP violating phase $\theta_{_3}$ of gluino affects the neutron EDM obviously.
Supposing $B_{_X}=300\;{\rm GeV},\;\lambda_{_1}=\lambda_{_2}=\lambda_{_3}=0.1,\;|\mu_{_X}|=2400\;{\rm GeV}$
and the CP violating phase of $\mu_{_X}$ is zero $(\theta_{_X}=0)$, we plot the neutron EDM $d_n$ varying with
the phase $\theta_{_3}$ in Fig.(\ref{theta3}), with the condition
$m_{_{\tilde{Q}_1}}=m_{_{\tilde{U}_1}}=m_{_{\tilde{D}_1}}=MQ$. When $MQ=1000\;{\rm GeV}$ and $\theta_{_3}=\pm\frac{\pi}{2}$,
the absolute value of the neutron EDM $d_{n}$ reaches the biggest value $1.04\times10^{-25}\;(e.cm)$, which is very close
to the present experiment upper limit. For the same value of $\theta_{_3}$, the absolute value of $d_n$ with $MQ=1200\;{\rm GeV}$
is smaller than that with $MQ=1000\;{\rm GeV}$.

%%%%%%%%%%%%%%%%%%%%%%%%%%%%%%%%%%%%%%%%%%%%%%%%%%%%%

\begin{figure}[h]
\setlength{\unitlength}{1mm}
\centering
\includegraphics[width=4.0in]{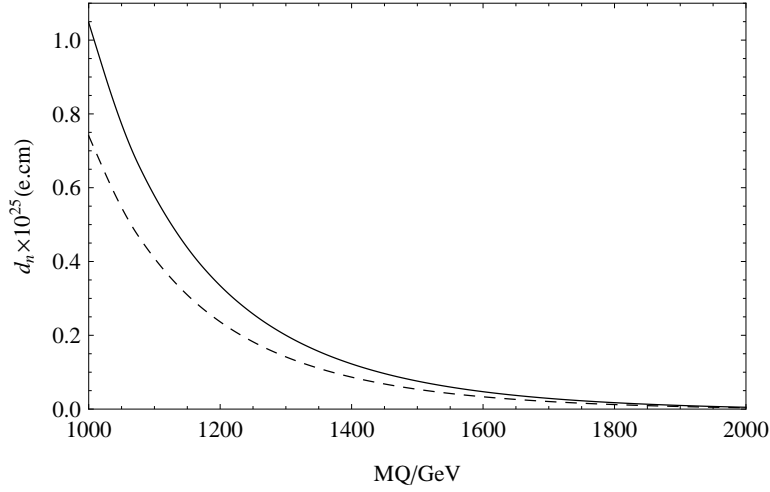}
\caption[]{As $B_{_X}=300\;{\rm GeV},\;\lambda_{_1}=\lambda_{_2}=\lambda_{_3}=0.1,\;|\mu_{_X}|=2400\;{\rm GeV},\;\theta_{_X}=0$
and $m_{_{\tilde{Q}_1}}=m_{_{\tilde{U}_1}}=m_{_{\tilde{D}_1}}=MQ$, the solid-line and dashed-line represent $d_n$ varying with $MQ$, for
$\theta_{_3}=\frac{\pi}{2}$ and $\frac{\pi}{4}$ respectively.   }\label{MQ}
\end{figure}

Then one finds $MQ$ affects the theoretical results strongly. Taking  $B_{_X}=300\;{\rm GeV},\;\lambda_{_1}=\lambda_{_2}=\lambda_{_3}=0.1,
\;|\mu_{_X}|=2400\;{\rm GeV}, \theta_{_X}=0,  \theta_{_3}=\frac{\pi}{2}(\frac{\pi}{4})$, and
$m_{_{\tilde{Q}_1}}=m_{_{\tilde{U}_1}}=m_{_{\tilde{D}_1}}=MQ$, the neutron EDM $d_n$ varying with
$MQ$ is plotted in Fig.(\ref{MQ}). Obviously, $d_n$ becomes small quickly, with the enlarging $MQ$ in the region
 from $1000\;{\rm GeV}$ to $1500\;{\rm GeV}$. When $MQ=2000\;{\rm GeV}$, $d_n$ turns very small, whose value is around
 $5\times10^{-28}\;(e.cm)$.

\begin{figure}[h]
\setlength{\unitlength}{1mm}
\centering
\includegraphics[width=4.0in]{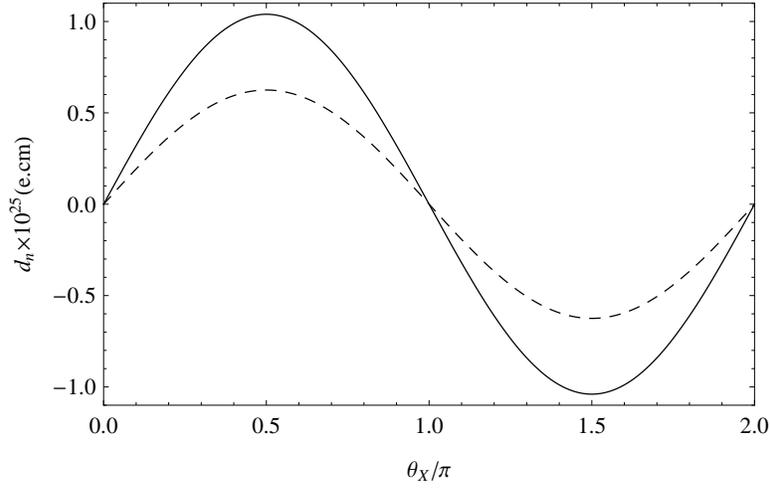}
\caption[]{As $B_{_X}=300\;{\rm GeV},\;\lambda_{_1}=\lambda_{_2}=\lambda_{_3}=0.1,\;\theta_{_3}=0$
and $m_{_{\tilde{Q}_1}}=m_{_{\tilde{U}_1}}=m_{_{\tilde{D}_1}}=1000\;{\rm GeV}$, the solid-line and dashed-line represent $d_n$ varying with $\theta_{_X}$, for
$|\mu_{_X}|=2400\;{\rm GeV}$ and $|\mu_{_X}|=3000\;{\rm GeV}$ respectively.   }\label{thetax}
\end{figure}

 The superfields X can give contributions to the neutron EDM $d_n$ at one-loop level, so the parameters having
 relation with the mass squared matrix of superfields X and the couplings of superfields X-exotic quark-quark can affect
 the theoretical predictions strongly. With the assumption in the parameter space $B_{_X}=300\;{\rm GeV},\;\lambda_{_1}=\lambda_{_2}=\lambda_{_3}=0.1,\;\theta_{_3}=0$
 and $m_{_{\tilde{Q}_1}}=m_{_{\tilde{U}_1}}=m_{_{\tilde{D}_1}}=1000\;{\rm GeV}$, we investigate the relation between $d_n$ and the
 CP violating phase $\theta_{_X}$ numerically in Fig.(\ref{thetax}). The shapes of the diagrams in Fig.(\ref{thetax}) are very similar as
 those in Fig.(\ref{theta3}). The biggest absolute value of $d_n$ is obtained with $\theta_{_X}=\frac{\pi}{2}$ and $|\mu_{_X}|=2400\;{\rm GeV}$.
 When $|\mu_{_X}|=3000\;{\rm GeV}$, the absolute value of $d_n$ is smaller than that with $|\mu_{_X}|=2400\;{\rm GeV}$.

\begin{figure}[h]
\setlength{\unitlength}{1mm}
\centering
\includegraphics[width=4.0in]{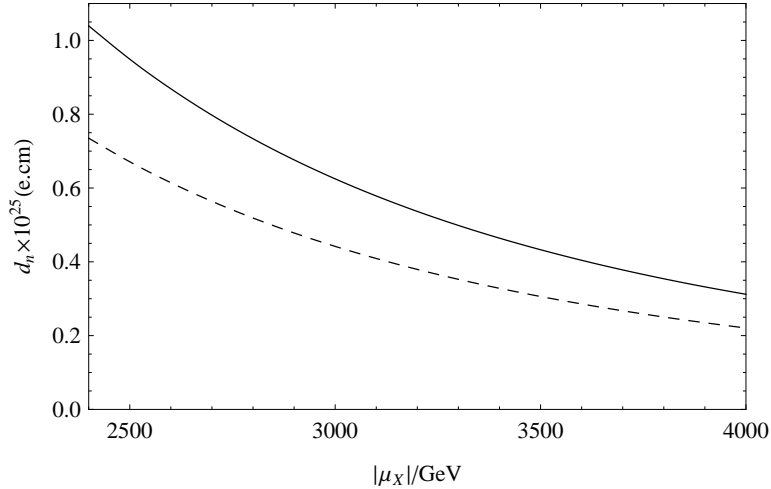}
\caption[]{As $B_{_X}=300\;{\rm GeV},\;\lambda_{_1}=\lambda_{_2}=\lambda_{_3}=0.1,\;\theta_{_3}=0$
and $m_{_{\tilde{Q}_1}}=m_{_{\tilde{U}_1}}=m_{_{\tilde{D}_1}}=1000\;{\rm GeV}$,
the solid-line and dashed-line represent $d_n$ varying with $|\mu_{_X}|$, for
$\theta_{_X}=\frac{\pi}{2}$ and $\frac{\pi}{4}$ respectively.   }\label{mux}
\end{figure}

In Fig.(\ref{mux}), we can easily find  the neutron EDM turns smaller and smaller, as the
absolute value of $\mu_{_X}$ becomes larger and larger.  The reason should be that some imaginary parts come from
the non-diagonal elements in the diagonalizing matrix $Z_{_X}$. Large diagonalized elements in the mass squared
matrix of the superfields X lead to small non-diagonal elements in $Z_{_X}$. Therefore, we obtain the small theoretical prediction
on the neutron EDM.

\begin{figure}[h]
\setlength{\unitlength}{1mm}
\centering
\includegraphics[width=4.0in]{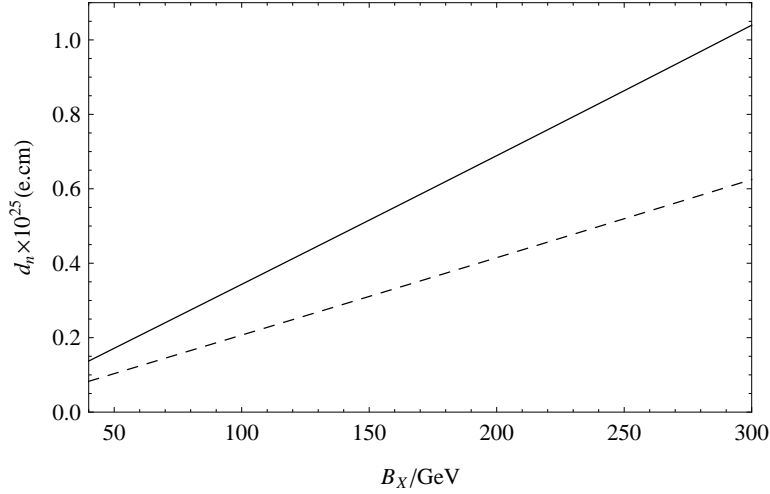}
\caption[]{As $\lambda_{_1}=\lambda_{_2}=\lambda_{_3}=0.1,\;\theta_{_3}=0,\;\theta_{_X}=\frac{\pi}{2}$
and $m_{_{\tilde{Q}_1}}=m_{_{\tilde{U}_1}}=m_{_{\tilde{D}_1}}=1000\;{\rm GeV}$, the solid-line and dashed-line represent $d_n$ varying with $B_{_X}$, for
$\;|\mu_{_X}|=2400\;{\rm GeV}$ and $3000{\rm GeV}$ respectively.   }\label{Bx}
\end{figure}

Choosing  $\lambda_{_1}=\lambda_{_2}=\lambda_{_3}=0.1,\;\theta_{_3}=0,\;\theta_{_X}=\frac{\pi}{2},\;
m_{_{\tilde{Q}_1}}=m_{_{\tilde{U}_1}}=m_{_{\tilde{D}_1}}=1000\;{\rm GeV}$, we plot $d_n$ varying
with the parameter $B_{_X}$ with $|\mu_{_X}|=2400(3000)\;{\rm GeV}$ in Fig.(\ref{Bx}).
From Fig.(\ref{Bx}), the neutron EDM $d_n$ is almost the linear increasing function of $B_{_X}$. Because $B_{_X}$
is the non-diagonal element in the mass squared
matrix of the superfields X, large $B_{_X}$ gives big contribution to imaginary elements in the matrix $Z_{_X}$.

\begin{figure}[h]
\setlength{\unitlength}{1mm}
\centering
\includegraphics[width=4.0in]{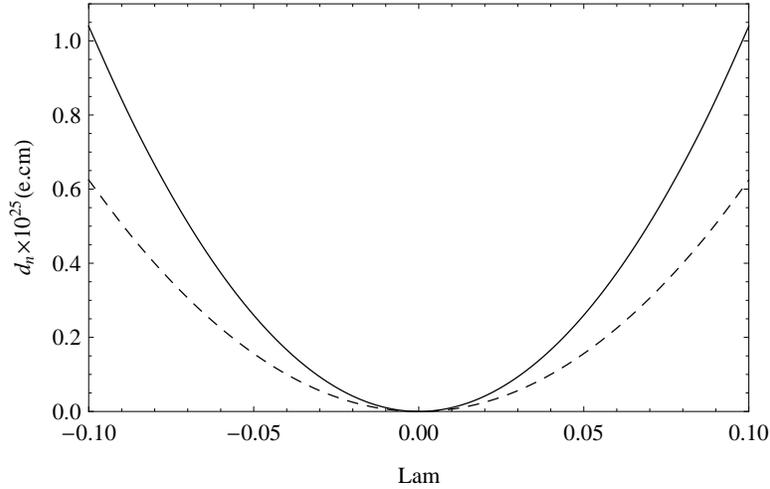}
\caption[]{As $B_{_X}=300\;{\rm GeV},\;\theta_{_3}=0,\;\theta_{_X}=\frac{\pi}{2},\;m_{_{\tilde{Q}_1}}
=m_{_{\tilde{U}_1}}=m_{_{\tilde{D}_1}}=1000\;{\rm GeV}$ and $\lambda_{_1}=\lambda_{_2}=\lambda_{_3}=Lam$,
the solid-line and dashed-line represent $d_n$ varying with $Lam$, for
$|\mu_{_X}|=2400{\rm GeV}$ and $3000{\rm GeV}$ respectively. } \label{Lam}
\end{figure}

From the Lagrangian of quark-exotic quark and the superfields X in Eqs.(\ref{LX}) and (\ref{couplingLX}),
one finds $\lambda_{_1},\;\lambda_{_2},\;\lambda_{_3}$
are important parameters, and they affect those coupling strongly. So, we analyse how $\lambda_{_1},\;\lambda_{_2},\;\lambda_{_3}$
influence the neutron EDM with the hypothesis in parameter space as $B_{_X}=300\;{\rm GeV},\;\theta_{_3}=0,\;\theta_{_X}=\frac{\pi}{2},\;m_{_{\tilde{Q}_1}}
=m_{_{\tilde{U}_1}}=m_{_{\tilde{D}_1}}=1000\;{\rm GeV}$ and $\lambda_{_1}=\lambda_{_2}=\lambda_{_3}=Lam$. The numerical
results are plotted in the Fig.(\ref{Lam}) for $\mu_{_X}=2400(3000)\;{\rm GeV}$.
When $Lam$ is very small, neutron EDM $d_{n}$ is almost zero. The diagrams in Fig.(\ref{Lam}) is symmetrical for the axis
$Lam=0$.

\section{summary \label{sec4}}
 We study the neutron EDM in the framework of the BLMSSM. The constraints from the experiment data on the
 lightest CP-even Higgs mass $(m_{h^0}=125.9\;{\rm GeV})$ and Higgs decay into diphoton $(ZZ^*,\;WW^*)$ are all considered. Furthermore
 the confine on the heaviest CP-even Higgs and CP-odd Higgs mass is also taking into account
 $(m_{_A^0},m_{_H^0})>600\;{\rm GeV}$. We analyse
 how the neutron EDM depends on the related parameters. If the new physics mass at TeV range,  the present experiment upper bound on the neutron EDM
 gives a serious constraint on the CP violating phase and the BLMSSM parameters.  Our numerical results show that in some
 parameter space, the neutron EDM $d_n$ can reach it's present experimental upper bound $1.1\times 10^{-25}(e.cm)$.

\appendix
\section{The couplings between Higgs and exotic quarks\label{app1}}
The coupling constants $(\mathcal{N}^L_{_{H^0_{_{\alpha}}}})(\alpha=1,2 \cdots 8)$ for
neutral Higgs and up type exotic quarks are expressed as
{\small\begin{eqnarray}
&&(\mathcal{N}^L_{_{H^0_{_{1}}}})_{_{ij}}=\frac{1}{\sqrt{2}}\Big[Y_{_{u_4}}(W_{_t}^\dagger)_{_{i2}}(U_{_t})_{_{1j}}\cos\alpha
+Y_{_{u_5}}(W_{_t}^\dagger)_{_{i1}}(U_{_t})_{_{2j}}\sin\alpha\Big],\nonumber\\&&
(\mathcal{N}^R_{_{H^0_{_{1}}}})_{_{ij}}=\frac{1}{\sqrt{2}}\Big[Y^*_{_{u_4}}(U_{_t}^\dagger)_{_{i1}}(W_{_t})_{_{2j}}\cos\alpha
+Y^*_{_{u_5}}(U_{_t}^\dagger)_{_{i2}}(W_{_t})_{_{1j}}\sin\alpha\Big],\nonumber\\&&
(\mathcal{N}^L_{_{H^0_{_{2}}}})_{_{ij}}=\frac{1}{\sqrt{2}}\Big[Y_{_{u_4}}(W_{_t}^\dagger)_{_{i2}}(U_{_t})_{_{1j}}\sin\alpha
-Y_{_{u_5}}(W_{_t}^\dagger)_{_{i1}}(U_{_t})_{_{2j}}\cos\alpha\Big],\nonumber\\&&
(\mathcal{N}^R_{_{H^0_{_{2}}}})_{_{ij}}=\frac{1}{\sqrt{2}}[Y^*_{_{u_4}}(U_{_t}^\dagger)_{_{i1}}(W_{_t})_{_{2j}}\sin\alpha
-Y^*_{_{u_5}}(U_{_t}^\dagger)_{_{i2}}(W_{_t})_{_{1j}}\cos\alpha\Big],\nonumber\\&&
%%%%%%%%%%%%%%%%%%%%%%%%%%%%%%%%%%%%%%%%%%%%%%%%%%%%%%%%%%%%%%%%%%%%%%%%%%
(\mathcal{N}^L_{_{H^0_{_{3}}}})_{_{ij}}=\frac{i}{\sqrt{2}}\Big[Y_{_{u_4}}(W_{_t}^\dagger)_{_{i2}}(U_{_t})_{_{1j}}\cos\beta
+Y_{_{u_5}}(W_{_t}^\dagger)_{_{i1}}(U_{_t})_{_{2j}}\sin\beta\Big],\nonumber\\&&
(\mathcal{N}^R_{_{H^0_{_{3}}}})_{_{ij}}=-\frac{i}{\sqrt{2}}\Big[Y^*_{_{u_4}}(U_{_t}^\dagger)_{_{i1}}(W_{_t})_{_{2j}}\cos\beta
+Y^*_{_{u_5}}(U_{_t}^\dagger)_{_{i2}}(W_{_t})_{_{1j}}\sin\beta\Big],\nonumber\\&&
(\mathcal{N}^L_{_{H^0_{_{4}}}})_{_{ij}}=\frac{i}{\sqrt{2}}\Big[Y_{_{u_4}}(W_{_t}^\dagger)_{_{i2}}(U_{_t})_{_{1j}}\sin\beta
-Y_{_{u_5}}(W_{_t}^\dagger)_{_{i1}}(U_{_t})_{_{2j}}\cos\beta\Big],\nonumber\\&&
(\mathcal{N}^R_{_{H^0_{_{4}}}})_{_{ij}}=-\frac{i}{\sqrt{2}}\Big[Y^*_{_{u_4}}(U_{_t}^\dagger)_{_{i1}}(W_{_t})_{_{2j}}\sin\beta
-Y^*_{_{u_5}}(U_{_t}^\dagger)_{_{i2}}(W_{_t})_{_{1j}}\cos\beta\Big],\nonumber\\&&
%%%%%%%%%%%%%%%%%%%%%%%%%%%%%%%%%%%%%%%%%%%%%%%%%%%%%%%%%%%%%%%%%%%%%%
(\mathcal{N}^L_{_{H^0_{_{5}}}})_{_{ij}}=-\frac{1}{\sqrt{2}}\Big[\lambda_{_u}(W_{_t}^\dagger)_{_{i2}}(U_{_t})_{_{2j}}\cos\alpha_{_B}
-\lambda_{_Q}(W_{_t}^\dagger)_{_{i1}}(U_{_t})_{_{1j}}\sin\alpha_{_B}\Big],\nonumber\\&&
(\mathcal{N}^R_{_{H^0_{_{5}}}})_{_{ij}}=-\frac{1}{\sqrt{2}}\Big[\lambda^*_{_u}(U_{_t}^\dagger)_{_{i2}}(W_{_t})_{_{2j}}\cos\alpha_{_B}
-\lambda^*_{_Q}(U_{_t}^\dagger)_{_{i2}}(W_{_t})_{_{2j}}\sin\alpha_{_B}\Big],\nonumber\\&&
(\mathcal{N}^L_{_{H^0_{_{6}}}})_{_{ij}}=-\frac{1}{\sqrt{2}}[\lambda_{_u}(W_{_t}^\dagger)_{_{i2}}(U_{_t})_{_{2j}}\sin\alpha_{_B}
+\lambda_{_Q}(W_{_t}^\dagger)_{_{i1}}(U_{_t})_{_{1j}}\cos\alpha_{_B}\Big],\nonumber\\&&
(\mathcal{N}^R_{_{H^0_{_{6}}}})_{_{ij}}=-\frac{1}{\sqrt{2}}[\lambda_{_u}^*(U_{_t}^\dagger)_{_{i2}}(W_{_t})_{_{2j}}\sin\alpha_{_B}
+\lambda^*_{_Q}(U_{_t}^\dagger)_{_{i2}}(W_{_t})_{_{2j}}\cos\alpha_{_B}\Big],\nonumber\\&&
%%%%%%%%%%%%%%%%%%%%%%%%%%%%%%%%%%%%%%%%%%%%%%%%%%%%%%%%%%%%%%%%%%%%%%%%
(\mathcal{N}^L_{_{H^0_{_{7}}}})_{_{ij}}=-\frac{i}{\sqrt{2}}\Big[\lambda_{_u}(W_{_t}^\dagger)_{_{i2}}(U_{_t})_{_{2j}}\cos\beta_{_B}
-\lambda_{_Q}(W_{_t}^\dagger)_{_{i1}}(U_{_t})_{_{1j}}\sin\beta_{_B}\Big],
\nonumber\\&&(\mathcal{N}^R_{_{H^0_{_{7}}}})_{_{ij}}=\frac{i}{\sqrt{2}}\Big[\lambda^*_{_u}(U_{_t}^\dagger)_{_{i2}}(W_{_t})_{_{2j}}\cos\beta_{_B}
-\lambda^*_{_Q}(U_{_t}^\dagger)_{_{i2}}(W_{_t})_{_{2j}}\sin\beta_{_B}\Big],\nonumber\\&&
(\mathcal{N}^L_{_{H^0_{_{8}}}})_{_{ij}}=-\frac{i}{\sqrt{2}}\Big[\lambda_{_u}(W_{_t}^\dagger)_{_{i2}}(U_{_t})_{_{2j}}\sin\beta_{_B}
+\lambda_{_Q}(W_{_t}^\dagger)_{_{i1}}(U_{_t})_{_{1j}}\cos\beta_{_B}\Big],\nonumber\\&&
(\mathcal{N}^R_{_{H^0_{_{8}}}})_{_{ij}}=\frac{i}{\sqrt{2}}\Big[\lambda^*_{_u}(U_{_t}^\dagger)_{_{i2}}(W_{_t})_{_{2j}}\sin\beta_{_B}
+\lambda^*_{_Q}(U_{_t}^\dagger)_{_{i2}}(W_{_t})_{_{2j}}\cos\beta_{_B}\Big].
\end{eqnarray}}

We show the coupling constants $(\mathcal{N}^L_{_{H^0_{_{\alpha}}}})(\alpha=1,2 \cdots 8)$ for
neutral Higgs and down type exotic quarks here.
{\small\begin{eqnarray}
&&(\mathcal{K}^L_{_{H^0_{_1}}})_{_{ij}}=\frac{1}{\sqrt{2}}\Big[Y_{_{d_4}}(W_{_b}^\dagger)_{_{i2}}(U_{_b})_{_{1j}}\sin\alpha
-Y_{_{d_5}}(W_{_b}^\dagger)_{_{i1}}(U_{_b})_{_{2j}}\cos\alpha\Big],\nonumber\\&&
(\mathcal{K}^R_{_{H^0_{_1}}})_{_{ij}}=\frac{1}{\sqrt{2}}[Y^*_{_{d_4}}(U_{_b}^\dagger)_{_{i1}}(W_{_b})_{_{2j}}\sin\alpha
-Y^*_{_{d_5}}(U_{_b}^\dagger)_{_{i2}}(W_{_b})_{_{1j}}\cos\alpha\Big],\nonumber\\&&
(\mathcal{K}^L_{_{H^0_{_2}}})_{_{ij}}=-\frac{1}{\sqrt{2}}\Big[Y_{_{d_4}}(W_{_b}^\dagger)_{_{i2}}(U_{_b})_{_{1j}}\cos\alpha
+Y_{_{d_5}}(W_{_b}^\dagger)_{_{i1}}(U_{_b})_{_{2j}}\sin\alpha\Big],\nonumber\\&&
(\mathcal{K}^R_{_{H^0_{_2}}})_{_{ij}}=-\frac{1}{\sqrt{2}}\Big[Y^*_{_{d_4}}(U_{_b}^\dagger)_{_{i1}}(W_{_b})_{_{2j}}\cos\alpha
+Y^*_{_{d_5}}(U_{_b}^\dagger)_{_{i2}}(W_{_b})_{_{1j}}\sin\alpha\Big],\nonumber\\&&
%%%%%%%%%%%%%%%%%%%%%%%%%%%%%%%%%%%%%%%%%%%%%%%%%%%%%%%%%%%
(\mathcal{K}^L_{_{H^0_{_3}}})_{_{ij}}=\frac{i}{\sqrt{2}}\Big[Y_{_{d_4}}(W_{_b}^\dagger)_{_{i2}}(U_{_b})_{_{1j}}\sin\beta
-Y_{_{d_5}}(W_{_b}^\dagger)_{_{i1}}(U_{_b})_{_{2j}}\cos\beta\Big],\nonumber\\&&
(\mathcal{K}^R_{_{H^0_{_3}}})_{_{ij}}=-\frac{i}{\sqrt{2}}\Big[Y^*_{_{d_4}}(U_{_b}^\dagger)_{_{i1}}(W_{_b})_{_{2j}}\sin\beta
-Y^*_{_{d_5}}(U_{_b}^\dagger)_{_{i2}}(W_{_b})_{_{1j}}\cos\beta\Big],\nonumber\\&&
(\mathcal{K}^L_{_{H^0_{_4}}})_{_{ij}}=-\frac{i}{\sqrt{2}}\Big[Y_{_{d_4}}(W_{_b}^\dagger)_{_{i2}}(U_{_b})_{_{1j}}\cos\beta
+Y_{_{d_5}}(W_{_b}^\dagger)_{_{i1}}(U_{_b})_{_{2j}}\sin\beta\Big],\nonumber\\&&
(\mathcal{K}^R_{_{H^0_{_4}}})_{_{ij}}=\frac{i}{\sqrt{2}}\Big[Y^*_{_{d_4}}(U_{_b}^\dagger)_{_{i1}}(W_{_b})_{_{2j}}\cos\beta
+Y^*_{_{d_5}}(U_{_b}^\dagger)_{_{i2}}(W_{_b})_{_{1j}}\sin\beta\Big],\nonumber\\&&
%%%%%%%%%%%%%%%%%%%%%%%%%%%%%%%%%%%%%%%%%%%%%%%%%%%%%%%%%%
(\mathcal{K}^L_{_{H^0_{_5}}})_{_{ij}}=-\frac{1}{\sqrt{2}}\Big[\lambda_{_d}(W_{_b}^\dagger)_{_{i2}}(U_{_b})_{_{2j}}\cos\alpha_{_B}
+\lambda_{_Q}(W_{_b}^\dagger)_{_{i1}}(U_{_b})_{_{1j}}\sin\alpha_{_B}\Big],\nonumber\\&&
(\mathcal{K}^R_{_{H^0_{_5}}})_{_{ij}}=-\frac{1}{\sqrt{2}}\Big[\lambda^*_{_d}(U_{_b}^\dagger)_{_{i2}}(W_{_b})_{_{2j}}\cos\alpha_{_B}
+\lambda^*_{_Q}(U_{_b}^\dagger)_{_{i2}}(W_{_b})_{_{2j}}\sin\alpha_{_B}\Big],\nonumber\\&&
(\mathcal{K}^L_{_{H^0_{_6}}})_{_{ij}}=-\frac{1}{\sqrt{2}}\Big[\lambda_{_d}(W_{_b}^\dagger)_{_{i2}}(U_{_b})_{_{2j}}\sin\alpha_{_B}
+\lambda_{_Q}(W_{_b}^\dagger)_{_{i1}}(U_{_b})_{_{1j}}\cos\alpha_{_B}\Big],\nonumber\\&&
(\mathcal{K}^R_{_{H^0_{_6}}})_{_{ij}}=-\frac{1}{\sqrt{2}}\Big[\lambda^*_{_d}(U_{_b}^\dagger)_{_{i2}}(W_{_b})_{_{2j}}\sin\alpha_{_B}
+\lambda^*_{_Q}(U_{_b}^\dagger)_{_{i2}}(W_{_b})_{_{2j}}\cos\alpha_{_B}\Big],\nonumber\\&&
%%%%%%%%%%%%%%%%%%%%%%%%%%%%%%%%%%%%%%%%%%%%%%%%%%%%%%%%
(\mathcal{K}^L_{_{H^0_{_7}}})_{_{ij}}=-\frac{i}{\sqrt{2}}\Big[\lambda_{_d}(W_{_b}^\dagger)_{_{i2}}(U_{_b})_{_{2j}}\cos\beta_{_B}
+\lambda_{_Q}(W_{_b}^\dagger)_{_{i1}}(U_{_b})_{_{1j}}\sin\beta_{_B}\Big],\nonumber\\&&
(\mathcal{K}^R_{_{H^0_{_7}}})_{_{ij}}=\frac{i}{\sqrt{2}}\Big[\lambda^*_{_d}(U_{_b}^\dagger)_{_{i2}}(W_{_b})_{_{2j}}\cos\beta_{_B}
+\lambda^*_{_Q}(U_{_b}^\dagger)_{_{i2}}(W_{_b})_{_{2j}}\sin\beta_{_B}\Big],\nonumber\\&&
(\mathcal{K}^L_{_{H^0_{_8}}})_{_{ij}}=-\frac{i}{\sqrt{2}}\Big[\lambda_{_d}(W_{_b}^\dagger)_{_{i2}}(U_{_b})_{_{2j}}\sin\beta_{_B}
+\lambda_{_Q}(W_{_b}^\dagger)_{_{i1}}(U_{_b})_{_{1j}}\cos\beta_{_B}\Big],\nonumber\\&&
(\mathcal{K}^R_{_{H^0_{_8}}})_{_{ij}}=\frac{i}{\sqrt{2}}\Big[\lambda^*_{_d}(U_{_b}^\dagger)_{_{i2}}(W_{_b})_{_{2j}}\sin\beta_{_B}
+\lambda^*_{_Q}(U_{_b}^\dagger)_{_{i2}}(W_{_b})_{_{2j}}\cos\beta_{_B}\Big].
\end{eqnarray}}

\begin{acknowledgments}

The work has been supported by the National Natural Science Foundation of China (NNSFC)
with Grant No. 11275036, No. 11047002, the Fund of Natural Science Foundation
of Hebei Province(A2011201118), the open project of State Key Laboratory of Chinese Academy of Sciences Institute
 of theoretical physics(Y3KF311CJ1)
 and Natural Science Fund of Hebei University
with Grant No. 2011JQ05, No. 2012-242.
\end{acknowledgments}

%%%%%%%%%%%%%%%%%%%%%%%%%%%%%%%%%%%%%%%%%%%%%%%%%%%%%%%%%%%%%%%%%%%
\end{document}